\documentclass[aps, amsmath, twocolumn, amssymb,groupedaddress]{revtex4}
\setcitestyle{super}
\usepackage{graphicx}
\usepackage{gensymb}
\pdfoutput=1
\usepackage{amsmath}
\usepackage{color}
\usepackage{float}
\usepackage{lipsum}
\usepackage{color,soul}

\begin{document}
\title{A modular point contact spectroscopy probe for sub-Kelvin applications}
\author{Shekhar Das}
\author{Goutam Sheet}
\email{goutam@iisermohali.ac.in}
\affiliation{Department of Physical Sciences, Indian Institute of Science Education and Research Mohali, Sector 81, S. A. S. Nagar, Manauli, PO: 140306, India}

\date{\today}
\begin{abstract} 

We present the construction and performance of a plug-n-play type point contact spectroscopy probe equipped with a piezo-driven coarse approach mechanism for sub-kelvin applications. A modular assembly has been built which can be placed in or taken out from a sub-Kelvin socket mounted inside a He3 cryostat (with a 7 Tesla magnet) using an external manipulator. A simple transfer process using a removable vertical manipulator combined with a load-lock chamber makes the exchange of the sample/tip rig very easy without breaking the vacuum insulation of the sample chamber. We have successfully performed point-contact spectroscopic studies on few elemental superconductors with low T$_{c}$. We have also used the probe to perform spectroscopic measurements on another low temperature superconductor PdTe$_2$, which is a Dirac semimetal in the normal state. We present such data to demonstrate the functionality of the probe. The probe can also be made adaptable for other transport measurements such as Hall effect, four probe resistivity, scanning tunnelling spectroscopy etc.
 
 \end{abstract}

\maketitle
{\section{Introduction}}

Point contact spectroscopy (PCS)\cite{naidyuk, daghero, janson, proslier} is a tool utilised to study the energy dependent differential conductance (and its higher orders) of a mesoscopic confined region and relies on the fact that I-V characteristics of a mesoscopic point contact between metals, show deviation from linear ohmic characteristics due to the interaction of the electrons with elementary excitation. This involves formation of mesoscopic point contacts between two materials and then studying the I-V characteristics of the contact region to obtain energy, momentum and spin resolved information about the Fermi surface.

 PCS has been established as one of the most powerful techniques for studying superconductors. In superconducting point contacts, the transport is dominated by a process called Andreev reflection\cite{andreev}. Measurement of Andreev reflection directly provides gap amplitude of the superconductor and its symmetry in the momentum space. This technique has widely been used to characterize  many elemental and complex superconductors such as, the heavy Fermion superconductors\cite{mikael}, BiS$_2$ based superconductors\cite{aslam1,aslam2}, pnictide superconductors\cite{green}, multi-gap superconductors\cite{multi} etc.  Besides these, the technique has also been employed to measure degree of spin-polarization in  ferromagnetic materials\cite{Soulen, gs, mazin_prl99} and topological insulators\cite{Borisov}. Using this thechnique, exotic quantum phenomena like quantum phase-slip, ballistic electron focusing, weak link Josephson effect\cite{josephson, rose} can also be realized. Very recently, the discovery\cite{leena_nm, taas, tci, majo1} of unexpected superconductivity in mesoscopic junctions of non-superconducting topological materials\cite{hasan_rmp, ando} and elemental metals has opened up a new paradigm in understanding the emergence of superconductivity in various materials. In this context, PCS has proved to be one of the most efficient spectroscopic tool employed to characterize such nano scale systems.\\
      The simplest way to realize a mesoscopic contact is to physically touch two meterials such that a narrow constriction is formed between the two. The size of the constriction (= diameter  when the cross-section is assumed circular) can range from few angstroms (\r {A}) to few hundred nanometers (nm). A contact with a diameter in the prescribed range is comparable with the characteristic length-scales \cite{datta} such as  mean free path,  de Broglie wavelength, coherence length \emph{etc.} of electrons  for the materials forming the junction. For such contacts the electrons in the contact region accelerate freely under the application of a potential across it and lose energy only upon resonantly exciting elementary excitations in the material, giving rise to non linearities in the $I-V$ characteristics of the contact. These non linearities can be analysed to derive quantitative information (energy, momentum and spin resolved) about the scattering mechanisms of electron with phonons, magnons and other elementary excitations present.  The non-linearities in the $I-V$ characteristics appear as some distinct spectral features in the differential resistance $(dV/dI)$ vs. applied dc voltage bias $(V)$ spectra. These spectra correspond to electron transport through a point contact and can be classified into different regimes on the basis of ratio of mean free length $'l'$ of the materials forming the constriction and the contact diameter $'a'$. The resistance of the point contact is described by the Wexler's formula \cite{wexler}:  $R_{PC} = \frac{2h/e^2}{(ak_F)^2} + \Gamma (l/a)\frac{\rho (T)}{2a}$, where $h$ is the Planck's constant, $e$ is the charge of an electron, $a$ is the point-contact diameter, $k_F$ is the Fermi wave vector, $\Gamma(l/a)$ is a slowly varying function varies 0.7 to 1, $l$ is electronic mean free path, $\rho$ is the bulk resistivity of the material, and $T$ is the temperature at the point-contact.  The formula has two terms, the first term describes contribution due to the Sharvin resistance($R_S$) and the second term corresponds to the Maxwell's resistance ($R_M$). For a contact, with diameter less than the elastic mean free length of electrons, contribution to resistance comes only from the Sharvin term in the Wexler formula and is termed as a ballistic point contact\cite{prb04}. Transport through such a point contact is free of scattering  and there is no dissipation of energy within the contact region. For a ballistic point contact between a metal and superconductor, the electronic transport is dominated by the Andreev reflection process, and provides basis for the Point Contact Andreev Reflection (PCAR) spectroscopy. Occurence of Andreev reflection appears as dips in the $dV/dI$ vs $V$ spectra for a point-contact.  On the other hand, if the contact diameter is greater than the inelastic mean free length of electrons then the transport is dominated by scattering mechanisms and energy is dissipated within the contact which leads to Joule heating of the contact. This type of a contact is known as a thermal point contact\cite{prb04} and contribution to resistance comes dominantly from the Maxwell's term. The spectra obtained for thermal contacts are recognised by appearance of critical current dominated peaks in the differential resistance spectra. In an intermediate regime, the features of both of the spectra corresponding to earlier discussed regimes can be seen i.e., the Andreev dips and critical current peaks are obtained simultaneously. The resistance is described by the Wexler's formula, where both the Sharvin and Maxwell resistance contribute to the overall resistance of the contact. This kind of contact known as intermediate point-contact.
        
 In this paper, we present a detailed description of instrumentation and implementation of a home-built point contact probe which is capable of operating  optimally at ultra low temperatures, under high vacuum and in presence of high magnetic fields along with a detailed discussion on measurement automation and data analysis. Our design comprises of a detachable probe head which can be inserted in any cryostat with some minor modifications. Here we will demonstrate the capability and performance of our home-built probe which is designed to work in a JANIS He3 cryostat where a temperature of $\sim$ 292 mK can be achieved.  Using this probe assembly, we have successfully conducted spectroscopic studies of superconductivity in several samples. Here we present the point-contact spectroscopic study of superconductivity in PdTe$_{2}$/Ag, Zr/PtIr, Sn/Ag point contacts down to 400 mK in high magnetic fields.



\vspace{1cm}
\section{Experimental Details}

\subsection{Point-contact}

\begin{figure}[h!]
\includegraphics[scale=.4]{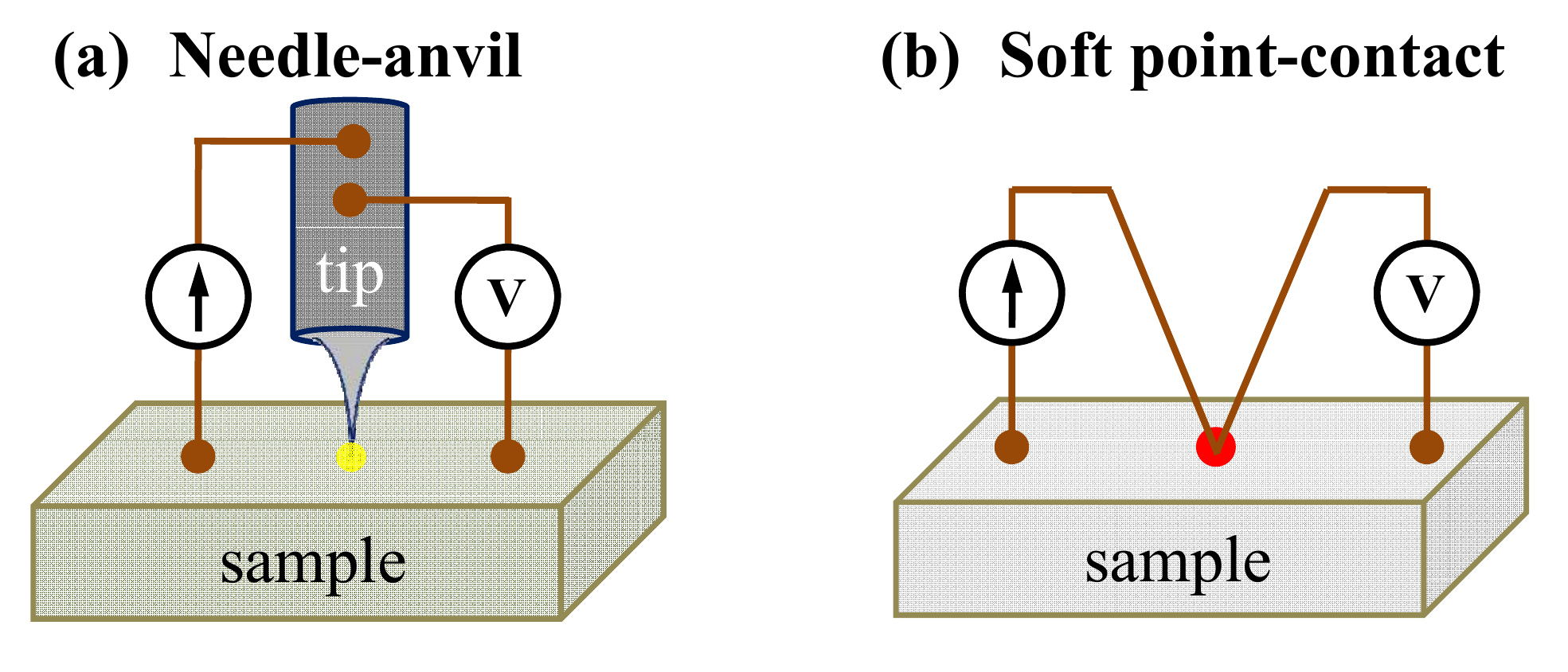}
\caption{Schematic of a point-contact set up with electrical connections. \textbf{(a)} The needle-anvil method: the point-contact is made by approaching a sharp tip (needle) towards the sample by manual differential screw or piezo-controlled nano-positioner. (\textbf{b)} The ``soft'' point-contact: Using small drop of Ag-paste is used to make a soft point-contact on the sample.}
\label{pc}
\end{figure}

The simplest way of creating a point contact is by touching a sharp metallic tip with a sample in a STM coarse approach like manner. This method is commonly known as the needle-anvil method where the tip serves as the needle\cite{naidyuk, daghero} (refer to Figure \ref{pc}). The tip movement is controlled either by use of a mechanical differential screw arrangement or by a piezo-controlled nano-positioner. Other methods\cite{naidyuk, daghero} of forming small constiction between materials include  shear method, break junction method, lithographic techniques and  ``soft'' contact method \cite{naidyuk,daghero} where the point contact is made by pasting a gold wire on a sample surface using a small drop of silver epoxy (EPO-TEK-EJ2189/1OZ, part A and Part B). For such soft contacts the effective cross-section of the point contact is tuned by application of current(or voltage) pulses through the contact. The pulses destroy some of the existing conduction channels and/or create new ones by piercing a thin oxide layer present on the surface of either electrode. The soft-contacts are mechanically and thermally  more stable than contacts formed by needle-anvil technique and does not involve pressure induced effects. However, the needle-anvil method  is non-destructive and provides better control over the contact diameter and is included in the probe design.\\

\subsection{Technical details}

In the present paper we discuss point contact spectroscopic measurements performed in a  liquid helium  He3 cryostat (Janis Research, USA). A He3 insert goes inside the bore of a superconducting solenoid capable of generating magnetic fields upto 7 Tesla in the vertical direction. A base temperature of 1.4 K is achieved in the insert by pumping over He4 liquid and is further cooled to 350 mK by pumping He3 liquid at a secondary stage.

\subsubsection{Design of the probe}
 
A measurement probe for a  He4 cooled system, which goes down to 1.4 K\cite{proslier,Tortello} and where exchange gas plays an important role to cool down the sample space, is quite easy to fabricate and handle.  
    \begin{figure*}[t]
\centering
\includegraphics[scale=.94]{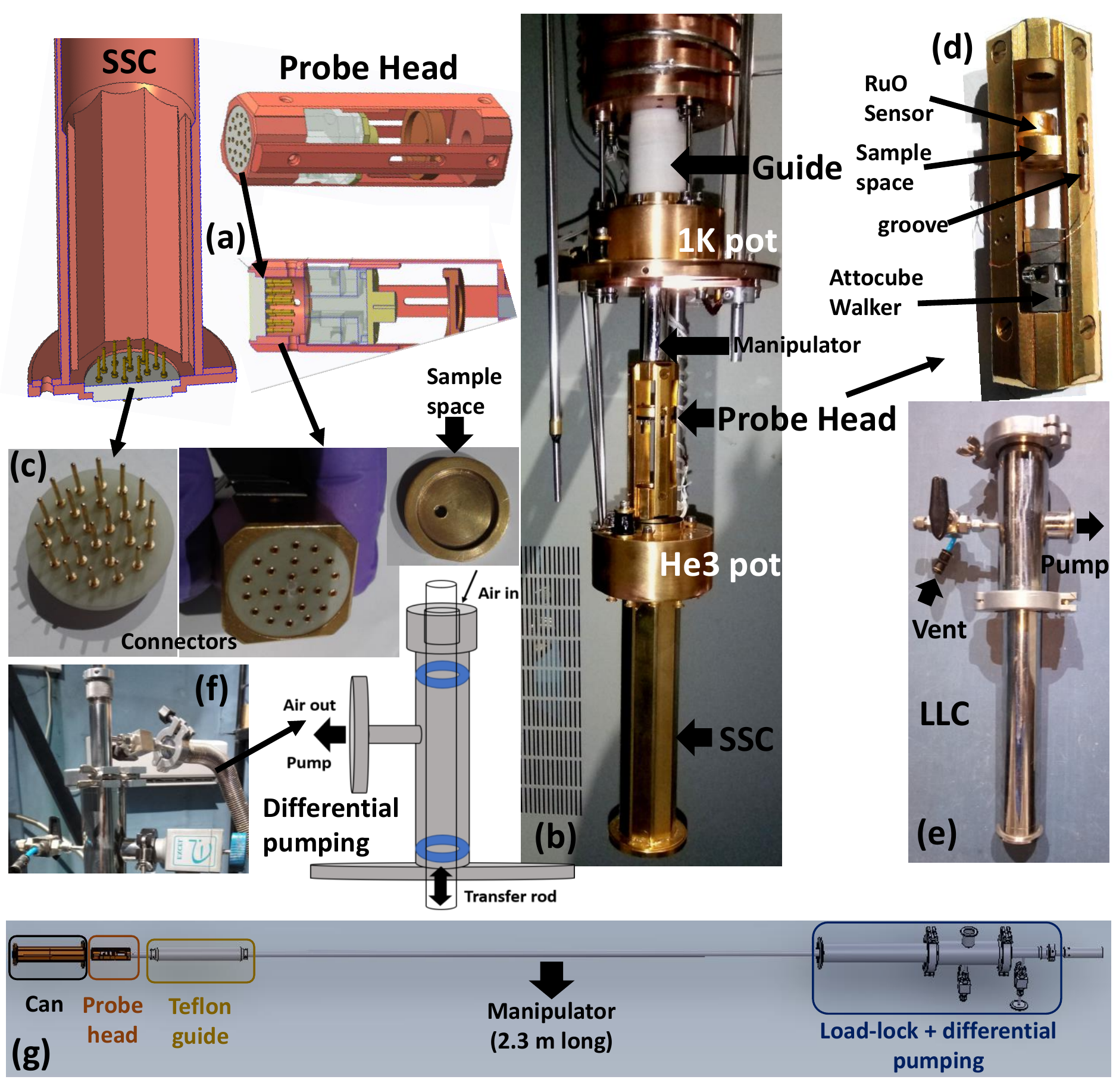}
\caption{Detachable probe-head (needle-anvil type) designed for sub-Kelvin operation. A transfer rod is used to place it inside the He-3 cryostat. The tip movement is controlled electronically through piezo-driven walkers. \textbf{(a)} Schematic of the head and sample space can. \textbf{(b)} Real image of the unit. \textbf{(c)} Different components of the probe-head. \textbf{(d)} Zoomed view of the probe head. \textbf{(e)} Load-lock chamber. \textbf{(f)} Differential pumping assembly. (g) Sample probe assembly.}
\label{piezo}
\end{figure*}
However, designing a probe for a He3 system, which goes down to sub-Kelvin temperature is a quite a challenging job as the sample probe needs to be transferred into a chamber maintained at high vacuum. The cooldown process of sample space in vacuum is even more challenging  where the cooling power decreases with the temperature and cooldown of probe assembly occurs via thermal contacts.  Hence it is essential to use material with high  thermal conductivity while keeping the thermal mass as low as possible. We have used oxygen-free high thermal conductivity (OFHC) grade copper to fabricate the probe and reduced the maximal amount of thermal mass to achieve the lowest possible temperature at the sample space. All of the probe components were gold coated to avoid any oxide layer formation which can be a major source of reduced/bad thermal conductivity. The PC probe setup is designed  to work optimally at ultra-low temperatures and high vacuum with an easy sample/tip changing process while the vacuum insulation of sample chamber remains undisturbed. \\
   The designed probe assembly comprises of four components, (1) a load-lock chamber (LLC) with a differential pumping mechanism  (figure 2(e,f)), (2) a vertical manipulator made of non-magnetic SS-316 (Figure 2(b)), (3) a probe head (Figure 2(d)), and (4) a sample space can (SSC) which is attached and thermally anchored with the coldest point of the cryostat (He-3 pot) (Figure 2(b)). The LLC is separated from an inner vacuum can (IVC) with a gate valve of size KF-40. The sample transfer mechanism is as follows: Firstly, the probe head (Sample space) is mounted on a threaded manipulator and is inserted into the LLC (Figure \ref{piezo} (e)). Then, the LLC is pumped down to $\sim$ 10$^{-6}$ mbar which is comparable to the pressure of IVC. Once the desired pressure is achieved in LLC, the probe head is inserted into the SSC with the help of the attached vertical  manipulator. A differential pumping mechanism(Figure \ref{piezo}(f)) has been employed to avoid any disruption in the vacuum of the IVC during probe insertion process. In this mechanism, the minimal amount of air which enters during the vertical movement of manipulator is trapped between two viton O-rings and pumped out with a roughing pump, thereby maintaining the vacuum pressure in IVC. Due to large length of the manipulator it gets difficult to align the rod to insert the probe into SSC hence, a teflon guide was used for maintaining horizontal alignment and is shown in Figure \ref{piezo}(b) \& (g). The diameter and length of guide is just right so that the probe head  aligns with the center of SSC where the mating grooves (made by using wire cutting method) present on the SSC and probe head helps in smooth insertion of the probe. Some alignment marks have been made on the manipulator to  correctly align these mating grooves. 
     The probe head contains a sample stage, a tip mount , piezo-driven motor(s), heater and sensor(s)(shown in Figure \ref{piezo}). All wire connections from these parts are connected to a 24 pin female mill-max connector arrangement embedded in a G-10 part (Figure \ref{piezo} (c)). This pin arrangement is fixed at the probe bottom using a good thermally conductive and electrically insulating epoxy glue (STYCAST 2850FT). The entire connector arrangement, along with the probe bottom, is well aligned so that it  exactly matches with the mating male part attached with the SSC bottom. All wires connections corresponding to the probe head parts enter into the cryostat top via feed-through and are connected to the male part at the bottom of SSC. After confirming that all the connections to the probe head are working properly, the probe head is un-threaded and  the manipulator is withdrawn to the LLC and gate valve is closed along with the thermal shutters. At this stage the sample is ready to be cooled.  A minimal amount (20 mbar approx.) of helium exchange gas is introduced into the IVC which begins cooling down the sample space by exchanging heat between the 1K pot and probe assembly. Once the sample space is cooled to 40 K the exchange gas is pumped out and a regular condensation procedure of He-3 is followed to achieve millikelvin temperature at the sample space (Details of the cooling process is described in the next paragraph). This entire process is completed in 4 to 5 hours after which experimentation on the sample is performed. Once the experiment  is over, the probe head is removed in a similar way. The probe head is never exposed to atmosphere directly after removing from the cold bath, and is kept in the LLC for a while to heat up to room temperature in presence of a minimal exchange gas. Here dry nitrogen gas is used for controlled venting of LLC and to protect the probe head from moisture.
         
\begin{figure}[h!]
\centering
\includegraphics[scale=0.8]{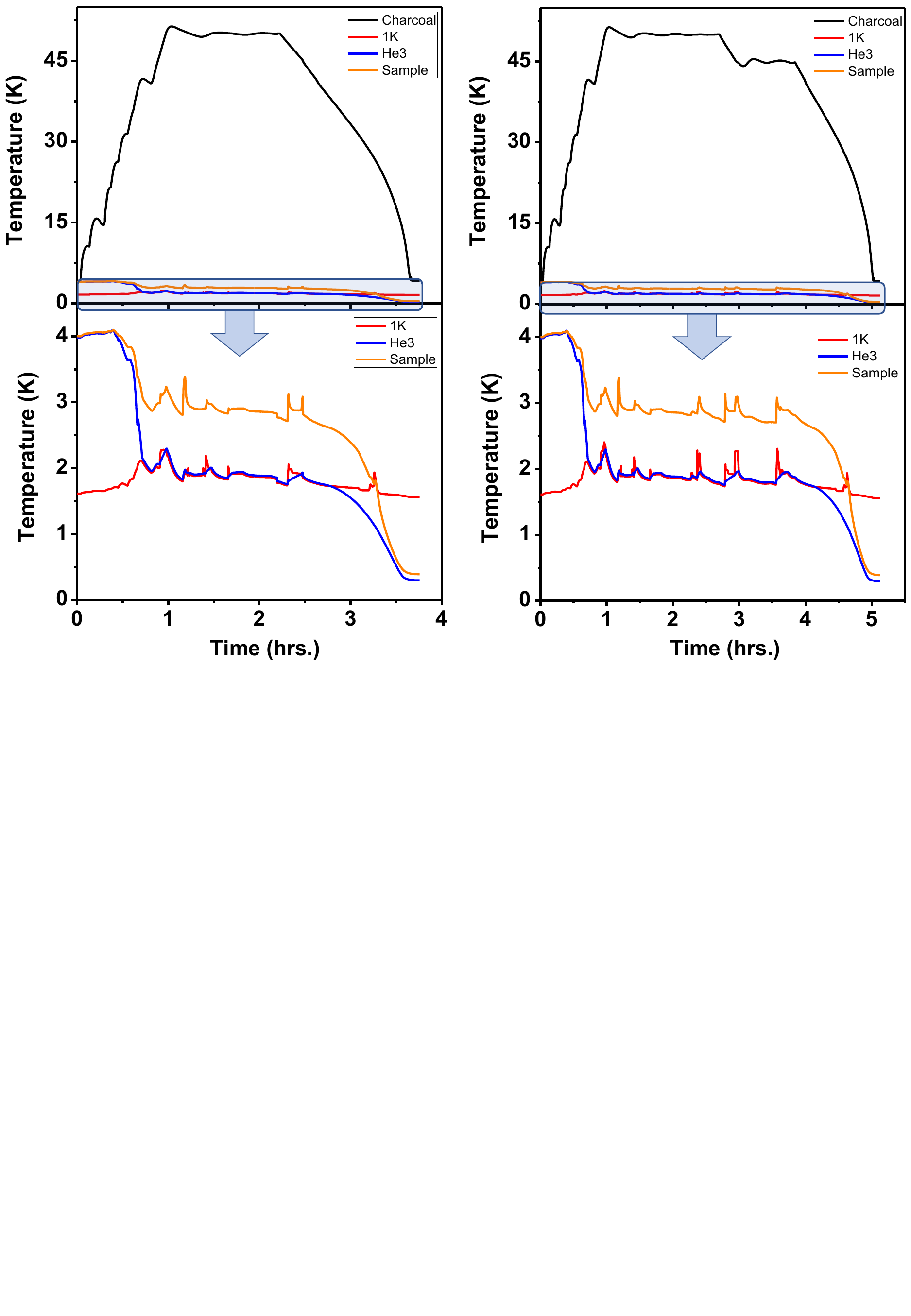}
\caption{(top) Temperature as a function of time during He3 condensation process. (bottom) A zoomed in view of the above graph.}
\label{Temp1}
\end{figure}

   While designing a probe one requires to take into account a number of factors. The most important of which is the cooling mechanism of sample space. A regular He3 condensation process is followed to achieve the base sub-kelvin temperature. In a regular condensation process, the temperature of charcoal (Figure \ref{Temp1}, black curve) is  gradually increased upto 50 K while maintaining the 1 K pot temperature (Figure \ref{Temp1}, red curve) below 2 K.  To maintain the  temperature of 1K pot  flow of liquid helium into the pot is controlled through a needle valve while pumping the pot by using a  dry roughing pump. A tube that connects the charcoal to the He3 pot, runs through the 1K pot and  He3 gas released from the charcoal condenses to form liquid (liquifaction temperature 3.2K) at 1K pot, drops down and is collected at the He3 pot. The collected liquid He3 is force pumped by cooling the charcoal pump and a temperature of about 292 mK(Figure \ref{Temp1}, blue curve) is achieved in the He3 pot. A representative cooling cycle is shown in Figure \ref{Temp1}. For the presented probe design, cooldown of the sample space  is achieved in multiple stages and wire connections need to be properly anchored for achieving sustainable cooldown to lower temperatures.  All the wire connections coming from the top of cryostat are thermally anchored with copper bobbins at multiple stages: (i) on the 4 K stage, (ii) on the 1 K pot, (iii) on the He3 pot. The heat-load dissipates in multiple stages and also cools down to the respective temperatures of the stages.  Manganin wires were used for the connections upto the connector on SSC and copper wires(0.25 mm diameter) were then used to distribute these connections further to increase the thermal conductivity.  The He3 pot is the coldest point of the insert and the SSC is attached to this pot by 12 gold-coated brass screws. These screws are non-magnetic and are thermally well connected with the He3 pot for improving thermal conduction. In addition to these screws, the He3 pot and SSC are also connected with high-quality copper braids. Since, the probe head is inserted into the SSC and remains in thermal contact with it, the cooldown of sample space occurs through thermal contact. The load and sample space cools down to a temperature of 347 mK (Figure \ref{Temp1}, orange curve). This base temperature holds for more than 48 hours if the 1 K pot temperature is maintained below 2 K. The difference between temperature of the He3 pot and sample space is primarily due to leakage of radiation from the top of the cryostat. The sample is glued on a copper sample-space using electrically insulating but thermally conducting G-varnish (sometimes thin quartz cover-slip is used in-between for better electrical insulation). Secondly, the sample is cooled by electrical wire connections present on the sample.

\begin{figure}[h!]
\centering
\includegraphics[scale=0.5]{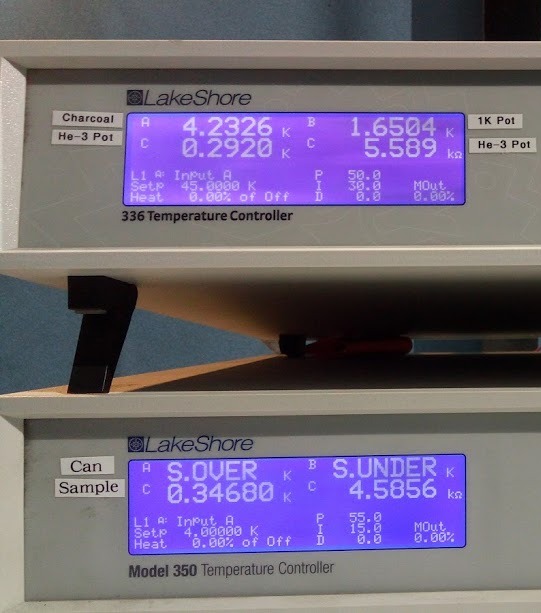}
\caption{Minimum temperature reading at He-3 pot (292 mK) and sample space (347 mK) using LakeShore temperature controller.}
\label{ff}
\end{figure}

The sample stage is equipped with a 50 $\ohm$ resistive heater and a calibrated ruthenium oxide temperature sensor. The tip assembly consists of a tip holder mounted on a piezo-driven three-stage (x-y-z) coarse posistioner (attocube ANPx101, ANPz101). The movement of the tip is electronically controlled through piezo-driven walkers similar to the coarse approach mechanism of STM and can be moved independently in all three directions. Power is supplied to these piezo positioners through low resistive wires (Cu-Ni coaxial). The sample space can also be moved up and down with respect to the tip holder depending on the sample and tip height. The metal tips can be fabricated either by electrochemical etching or simply by cutting metal wires (0.25\, mm in diameter). Two contact leads are fabricated on the tip and two more contacts are fabricated on the sample using silver-epoxy. Of these four electrodes, two are used to provide current and other two to sense the potential drop between the tip and sample. The probe has twelve free connections which allows us to simultaneously preform experiments on multiple samples. These connections also allows the user to perform several different experiments in the same probe assembly such as Hall effect, four probe resistivity etc.\\

\subsubsection{Electrical measurement and automation} 

 The point contact spectrum is obtained through lock-in based modulation technique. Current is provided to the point contact and corresponding voltage drop is recorded to obtain a point contact spectrum and corresponding analysis provides information about electron transport. A dc current $I_{dc}$  (Keithley 6221) coupled with a fixed, small amplitude ac current $I_{ac}cos\omega t$( SR830 Stanford Research Systems) is supplied to the point contact. The dc output voltage across the point contact is recorded by a digital multimeter (Keithley 2000) while the differential resistance $dV/dI$ (proportional to the ac output voltage locked internally to the first harmonic) is measured by the lock-in amplifier. The signal locked at first harmonic is proportional to $dV/dI$ and second harmonic to d$^2$V/dI$^2$. This can be shown by expressing the output voltage via Taylor expansion,\\ 
 \ $V(I=I_{dc}+I_{ac}cos\omega t$ with $I_{dc}\gg I_{ac}$)
$=V(I_{dc})+\left(\frac{dV}{dI}\right)\mid_{I_{dc}}I_{ac}cos\omega t+\frac{1}{2}\left(\frac{d^2V}{dI^2}\right)\mid_{I_{dc}}(I_{ac}cos\omega t)^2+\cdots=V(I_{dc})+\left(\frac{dV}{dI}\right)\mid_{I_{dc}}I_{ac}cos\omega t+\frac{1}{4}\left(\frac{d^2V}{dI^2}\right)\mid_{I_{dc}}(I_{ac})^2(1+cos2\omega t)+\cdots$ 

\begin{figure}[h!]
\centering
\includegraphics[scale=0.6]{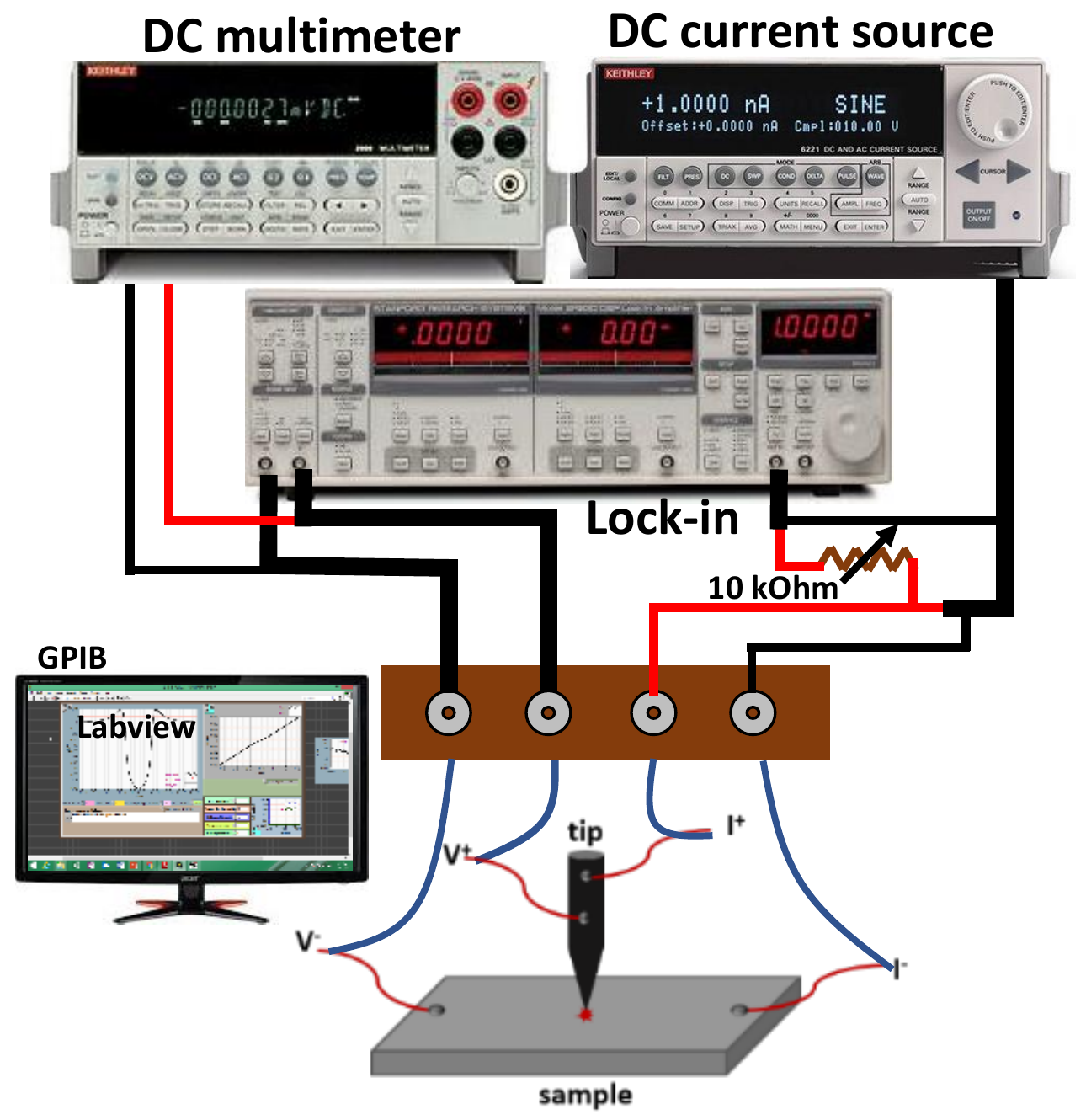}
\caption{Schematic of PCS measurement using lock-in based modulation technique.}
\label{EXPT_PCDA}
\end{figure}

\begin{figure}[h!]
\centering
\includegraphics[scale=0.52]{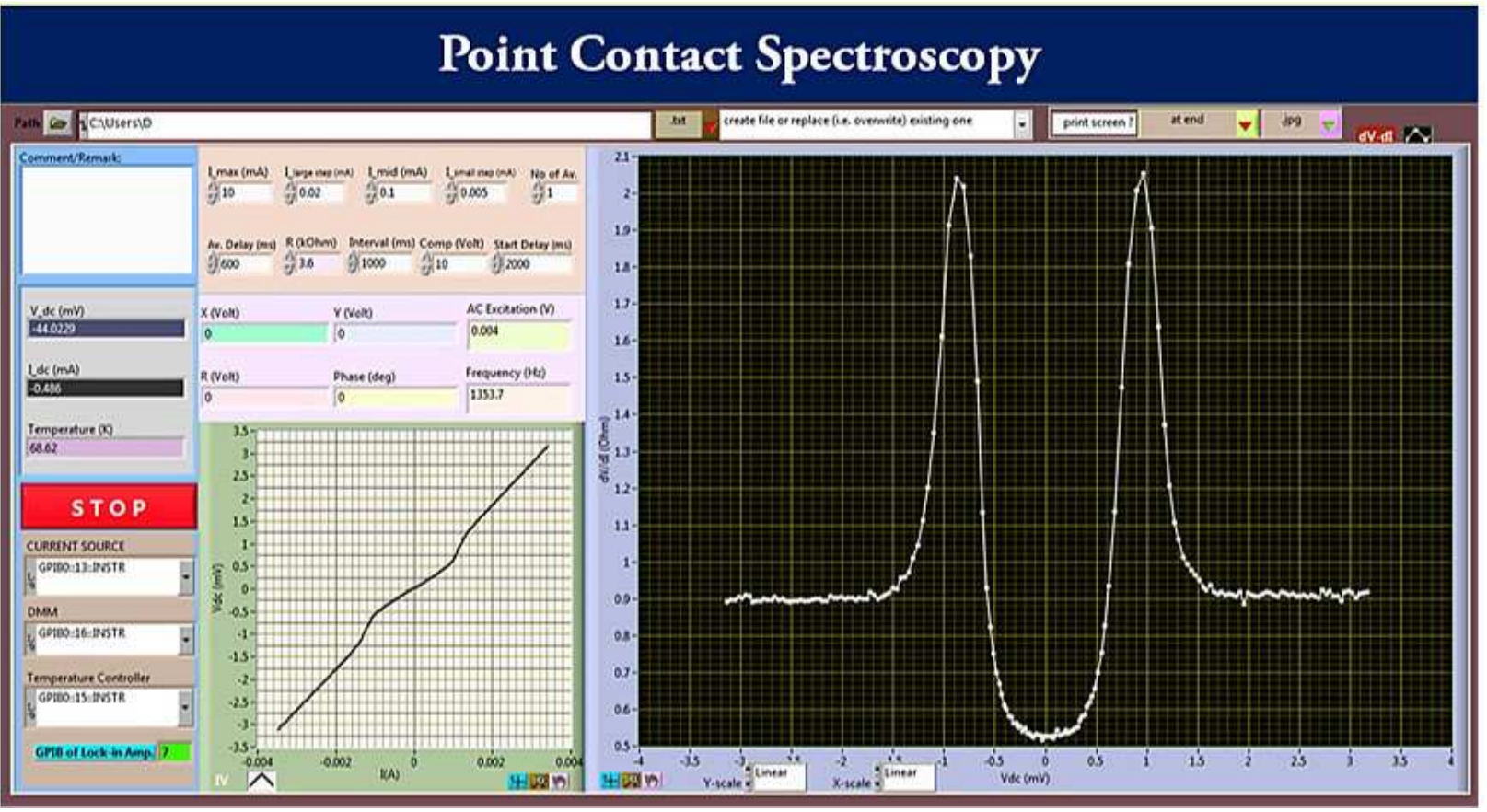}
\caption{Interactive and user-friendly window for automated measurement.}
\label{pcs11}
\end{figure}

One can also use a voltage-to-current converter with an adder circuit to couple dc voltage (from a source meter or from auxiliary port of SR830) and ac voltage (from SR830 with locked internally or from a function generator locked externally). The schematic of this technique is shown in Figure \ref{EXPT_PCDA}. An alternative measurement for the differential conductance can be performed in pulsed technique using Keithley 6221 with Keithley 2182A (nano volt meter). The pulsed technique measurement is very fast and suitable for low-power devices and/or in ultra-low temperature range (short pulses drastically reduces the Joule heating). The data is acquired through a GPIB connection and the measurement process is completely automated (using LabVIEW platform) with full control over the measurement parameters like temperature, magnetic field, \emph{etc.}, controlled through an interactive graphic-user-interface (GUI) as shown in Figure \ref{pcs11}.
 \\ 

\subsection{Data analysis software}
\begin{figure}[h!]
\centering
\includegraphics[scale=.7]{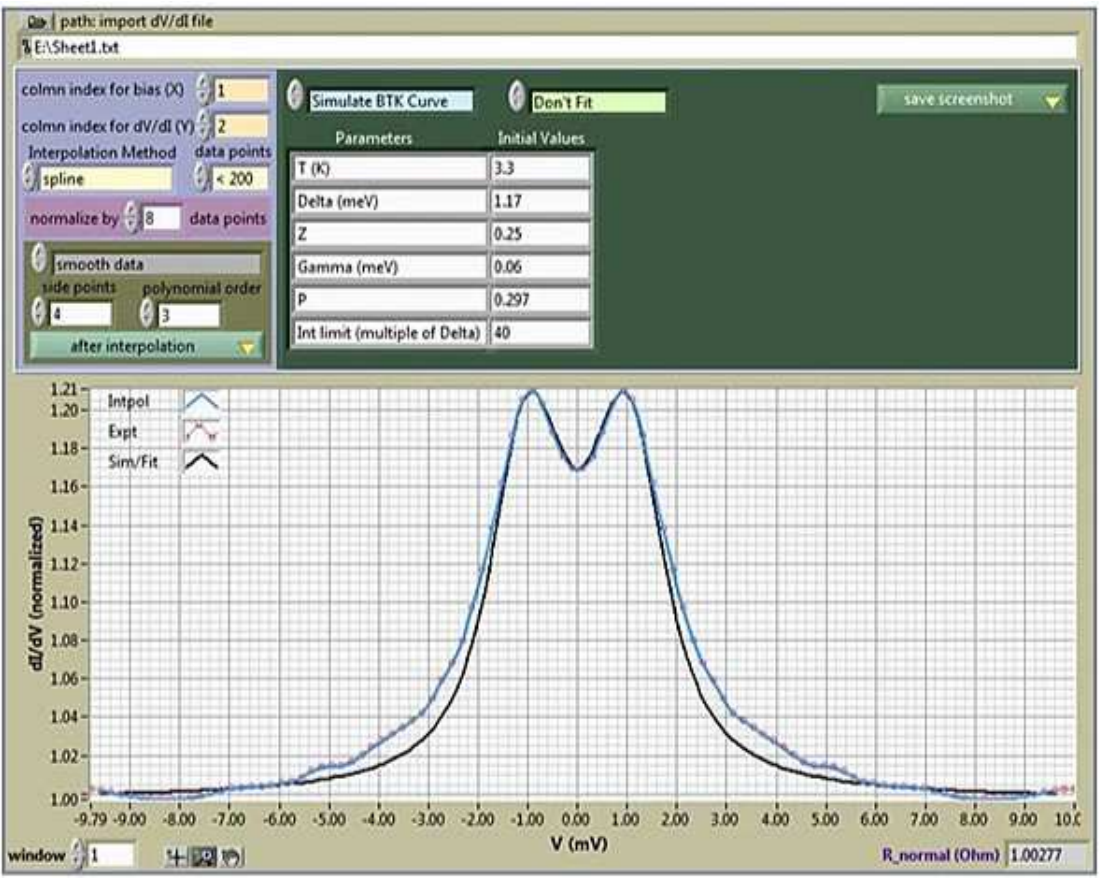}
\caption{Interactive window for data analysis. One can simulate/fit PCS spectrum following BTK theory.}
\label{pcs_sim}
\end{figure}

A LabVIEW based  interactive, user-friendly graphical user interface has been developed to obtain and analyze experimental data (refer to Fig. \ref{pcs_sim}). An interactive simulation program of PCS spectrum based on the Blonder-Tinkham-Klapwijk (BTK)\cite{btk} has also been developed where simulated spectrum can be obtained for a given superonducting gap ampliture ($\Delta$), temperature of the junction $(T)$, interface barrier strength $(Z)$, quasi-particle life-time ($\Gamma$)\cite{plecenik} and spin polarization $(P)$ and can model fit experimental spectra.


%
\section{Results and discussion}
\begin{figure*}[t]
\centering
\includegraphics[scale=.94]{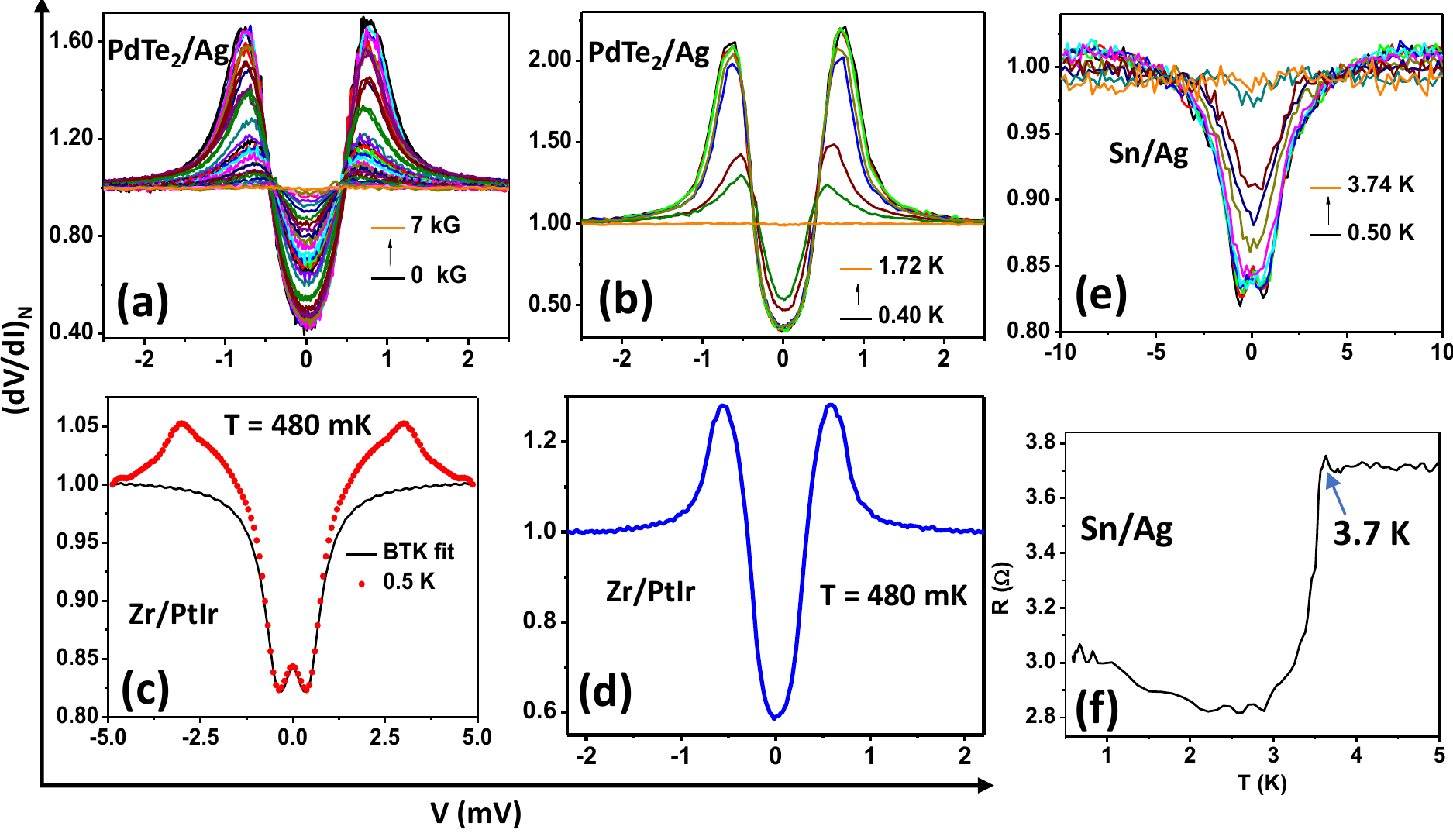}
\caption{(a) Magnetic field (b) temperature dependence of a PdTe$_{2}$/Ag PC  (c) A representative spectrum (red) of a Zr/PtIr PC with the BTK fit (black). (d) A representative thermal regime spectrum of a Zr/PtIr PC. (e) Temperature dependence of a Sn/Ag PC. (f) Superconducting transition at 3.7 K of a Sn/Ag PC.}
\label{expt}
\end{figure*}

The performance of the probe at low temperatures was guaged through several experiments. A point-contact study on PdTe${_2}$ was performed using a silver(Ag) tip.To check the performance of the probe, we performed a few experiments at low temperatures, key results of two such experiments  are provided here.  We carried out point-contact study on  PdTe$_{2}$ using Ag tip. PdTe$_{2}$ is a known superconductor with a critical temperature (T$_{c}$) of 1.7\cite{Gugg,prb}.  Point contact spectra obtained in the thermal regime of transport are presented in Figure \ref{expt}.  Spectral features obtained for superconducting PdTe$_{2}$  must evolve with the magnetic field and the temperature. Magnetic field dependence of a spectrum of PdTe$_{2}$/Ag PC is shown in Figure \ref{expt} (a). The spectral features evolve gradually and vanish at 7 kG. As expected the spectral features also evolved  with temperature and vanished at 
1.72 K (Figure \ref{expt} (b)) which is the bulk T$_{c}$ of PdTe$_{2}$ as per previous reports.
We also performed experiments on Zr/PtIr point-contacts. In Figure \ref{expt} (c), we present a spectrum obtained for intermediate regime of transport (red in color) along with the BTK simulated curve (black in color). The experimental data fits exactly with the simulated curve for bias values close to zero. The dips in the differential resistance about $V = 0$ is known as Andreev reflection dip.  The corresponding superconducting energy gap  is found to be $\Delta$ = 0.510 meV.

In Figure \ref{expt} (d) we have shown a representative spectrum of a thermal regime spectrum of Zr/PtIr point-contact where the peaks  correspond to critical current and are symmetric about $V = 0$.

Tin is a superconductor with T$_{c}$ of 3.7 K. We performed point-contact measurements on Sn with Ag tip. A temperature evolution of point-contact spectra is shown in Figure \ref{expt} (e). The bulk T$_{c}$ is also measured in point-contact geometry shown in Figure \ref{expt} (f). 
\\\\\\
\section{Conclusions}
In conclusion, we have  designed, fabricated, and installed a user-friendly probe assembly which is capable of operating down to sub-Kelvin temperatures. Utilizing this probe assembly point-contact measurements on PdTe$_2$, Zr and Sn at different magnetic fields and temperatures were successfully carried out.  The designed probe is fully capable to carry out various electronic transport measurements and enables the user to carry out STM/STS experiments with only a few minor modifications.

\section*{Acknowledgements}
We thank Mohammad Aslam and Sirshendu Gayen for their assistance  during various stages of this work. We thank Subash Pai and Mukesh Kamboj for their help in fabrication of the probe. GS acknowledges financial support from an SERB grant (\textbf{EMR/2015/001650}), for the construction of the probe.

\end{document}